\documentclass[cits]{PoS}

\title{NLTE abundances of Cr in the Sun and metal-poor stars}

\ShortTitle{NLTE abundances of Cr in the Sun and metal-poor stars}

\author{\speaker{Maria Bergemann}%
         \thanks{Based on observations made with the European Southern
Observatory telescopes (obtained from the ESO/ST-ECF Science Archive Facility)
and the Calar Alto Observatory telescopes.}\\
        Max-Planck-Institut f\"ur Astrophysik\\
        E-mail: \email{mbergema@mpa-garching.mpg.de}}

\abstract{We investigate statistical equilibrium of Cr in the atmospheres of
late-type stars. The main goal is to ascertain the reason for a systematic
abundance discrepancy between Cr I and Cr II lines, which is often encountered
in spectroscopic analyses of metal-poor stars. Up to now, all these studies
relied on the assumption of local thermodynamic equilibrium (LTE) in the
spectrum modelling. For the first time, we perform NLTE calculations in
subdwarfs and subgiants of different metallicities. We show that the LTE
assumption is inadequate to describe excitation-ionization equilibrium of Cr
I/Cr II in stellar atmospheres and, as a result, leads to large errors in
abundances. In particular, the NLTE abundance corrections to Cr I lines range
from $+0.3$ to $+0.5$ dex at low [Fe/H]. The NLTE [Cr/Fe] trend in the halo and
the disk is flat and can be reproduced by most of the models of Galactic
chemical evolution with standard prescriptions for Cr and Fe nucleosynthesis.}

\FullConference{11th Symposium on Nuclei in the Cosmos\\
                 19-23 July 2010 \\
                 Heidelberg, Germany.}

\begin{document}

\section{Introduction}

There is a well-known problem with modelling excitation and ionization balance
of Cr in the atmospheres of late-type stars. Systematic differences of $0.1 -
0.5$ dex between abundances based on LTE fitting of the Cr I and Cr II lines
were reported for metal-poor giants and dwarfs (e.g. Johnson 2002, Lai et al.
2008, Bonifacio et al. 2009). The discrepancies are smaller in the atmospheres
with larger metal content, amounting to $\sim0.1$ dex for Galactic disk stars
(Prochaska et al. 2000) and for the Sun (Sobeck et al. 2007, Asplund et al.
2009). It is common to attribute these offsets to the \emph{overionization} of
neutral Cr, a typical NLTE phenomenon affecting minority atoms in stellar
atmospheres. However, this explanation lacks theoretical justification, because
calculations of statistical equilibrium of Cr in model atmospheres under
restriction of different stellar parameters have not been performed up to now.

This discrepancy has important implications for Galactic chemical evolution
studies, which rely on gradients of observed abundance ratios. Depending on the
ionization stage of Cr used for abundance calculations in metal-poor stars,
radically different trends of [Cr/Fe] with [Fe/H] are obtained. The constant
[Cr/Fe] with metallicity, as derived from LTE analyses of Cr II lines, has a
simple interpretation in the theory of nucleosynthesis. Cr is co-produced with
Fe in explosive Si-burning that occurs in SNe, and the production ratio Cr/Fe is
roughly solar in both SNe II and SNe Ia. On the other side, there is no simple
explanation for declining [Cr/Fe] ratios with decreasing [Fe/H], which follows
from LTE analyses of Cr I lines. Recent studies of metal-free massive stars and
their nucleosynthesis yields show that \emph{subsolar} Cr/Fe abundance ratios in
very metal-poor stars can not be reproduced by any combination of SN II model
parameters (e.g. Heger \& Woosley 2008), especially when other Fe-peak elements
are taken into account.

Here, we report NLTE abundances of Cr for the Sun and a sample of dwarfs and
subgiants with $-3.2 \leq$ [Fe/H] $\leq -0.5$. The description of the methods is
given in Sect. \ref{sec:methods}. The statistical equilibrium of Cr under
restriction of different stellar parameters is discussed in Sect.
\ref{sec:stat}. The NLTE abundances for the observed stellar sample are
presented in Sect. \ref{sec:abundances}.

\section{Methods}{\label{sec:methods}}

Restricted NLTE calculations for Cr were performed with the code DETAIL and the
model of Cr atom was constructed with the data from the Kurucz
database\footnote{http://kurucz.harvard.edu/}. In short, the model includes
$339$ levels of Cr I/II and the number of radiatively-allowed transitions is
$6806$. The quantum-mechanical photoionization cross-sections for
quintet and septet states of Cr I were taken from Nahar (2009). For all other
levels, the Kramer's formula was adopted. For the rates of electronic and
inter-atomic collisions, we used standard prescriptions. In particular, we
investigated the sensitivity of results to the poorly-known rates of transitions
due to collisions with H I, which are usually approximated by the formulae of
Drawin (1969). The details about NLTE calculations, as well as parameters of the
lines selected for the abundance analysis, can be found in Bergemann \& Cescutti
(2010). The abundances of Cr were computed by a method of spectrum synthesis
with the code SIU (T. Gehren, private communication). All calculations were
performed with 1D plane-parallel models MAFAGS-ODF (Grupp 2004) with Kurucz's
opacity distribution functions. Stellar parameters, $T_{\rm eff}$, $\log g$, and
[Fe/H], were taken from Gehren et al. (2004, 2006), who used the same model
atmospheres.

\section{Statistical equilibrium of Cr}{\label{sec:stat}}

Similar to other complex minority atoms in the atmospheres of late-type
stars, such as Fe I, the overall distribution of atomic level populations
in Cr I is determined by the interplay of several processes. Overionization and
photon pumping are due to the non-local UV radiation field with mean intensity
larger than the local Planck function $J_{\rm \nu} > B_{\nu}(T_{\rm e})$. These
are counteracted by photon suction and overrecombination due to $J_{\rm \nu} <
B_{\nu}(T_{\rm e})$ at infra-red frequencies. Deviations from LTE caused by
non-thermal radiation field are balanced by collisional coupling between levels.
The relative role of these processes is different for stellar atmospheres with
different $T_{\rm eff}$, $\log g$, and [Fe/H].

In general, deviations from LTE in Cr I develop in the atmospheric layers where
the mean intensity exceeds the Planck function over the bound-free edges of
well-populated Cr I levels with excitation energies $\sim 2 - 4$ eV.
Overionization is particularly strong from the levels with large
quantum-mechanical photoionization cross-sections. For the Sun, the role of 
transitions in strong near-UV lines is non-negligible. They influence the Cr
I excitation balance in the outer layers of the solar model atmosphere, $\log
\tau_{\rm 500} < -2$. In the deeper regions, where many weak observable Cr I
lines are formed, the collisional interaction of the Cr I levels with each other
is quite strong. So that NLTE abundance corrections are relatively small, $\leq
0.1$ dex for the majority of lines.

In the atmospheres of cool metal-poor dwarfs and subgiants, radiative
processes dominate over collisions. NLTE effects on the levels of Cr I and Cr II
are amplified compared to the solar case and grow with decreasing [Fe/H].
As a result of low metal abundances, the effect of line-blanketing is reduced
leading to increased UV fluxes and collisional coupling between the levels is
very weak due to deficient electrons.
Deviations from LTE in Cr I are sensitive to effective temperature and gravity
at low metallicity. In the cool models, the effect of gravity is very
pronounced. For example, at [Fe/H] $= -3$, departures from LTE for Cr I levels
are stronger in the model with $T_{\rm eff} = 5000$ K and $\log g = 2.6$ than in
the model with $T_{\rm eff} = 6000$ K and $\log g = 4.2$, despite larger $T_{\rm
eff}$ of the latter model. This may account for a systematic difference between
metal-poor giants and dwarfs found by Lai et al. (2008) and Bonifacio et al.
(2009).

\section{Abundances of Cr for the Sun and metal-poor
stars}{\label{sec:abundances}}

The solar abundance was determined by comparing synthetic Cr lines with the
Solar Flux Atlas of Kurucz et al. (1984). Oscillator strengths for the
majority of Cr I transitions were taken from Sobeck et al. (2007), who estimate
the accuracy of $\log gf$'s to be better than $\sim 10\%$.

The solar LTE abundance of Cr determined from the Cr I lines is $5.66$ dex with
$\sigma = 0.04$ dex. This value is consistent with the meteoritic abundance,
$5.63 \pm 0.01$ dex\footnote{The Cr abundance in CI-chondrites from Lodders et
al. 2009 was renormalized to the photospheric Si abundance of Shi et al. (2008),
$\log\varepsilon_{\rm Si, \odot} = 7.52$ dex}. The LTE abundance based on the Cr
II lines is $5.81 \pm 0.13$ dex that makes a large abundance discrepancy between
the Cr I and Cr II lines, $\sim 0.15$ dex.
In contrast to LTE, the difference between both ionization stages in NLTE is
fairly small. Neglecting inelastic collision with H I in the calculations of Cr
population densities, $S_{\rm H} = 0$, we derive  $\log\epsilon = 5.74 \pm 0.05$
from the solar Cr I lines and $\log\epsilon = 5.79 \pm 0.12$ dex from the Cr II
lines. The results for $S_{\rm H} = 0.05$ are not very different from the
previous case, $\log\epsilon = 5.7 \pm 0.04$ dex (Cr I) and $\log\epsilon = 5.79
\pm 0.12$ dex (Cr II). A few Cr I and Cr II lines give systematically higher
abundances. Since these lines are sensitive to microturbulence parameter, this 
anomaly most likely reflects the shortcomings of our 1D model atmospheres.
\begin{figure}
\begin{center}
\includegraphics[clip,angle=0,width=.7\textwidth]{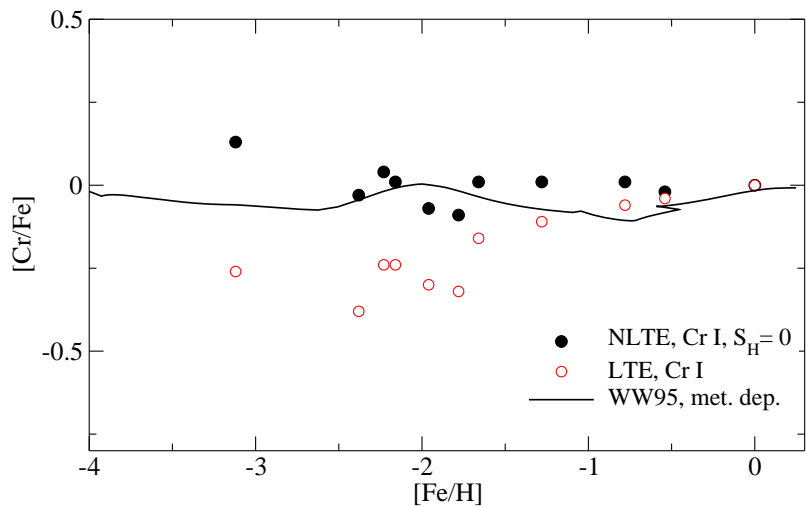}
\caption{Abundance ratios [Cr/Fe] as a function of metallicity. NLTE and
LTE-based Cr abundances in metal-poor stars are marked with filled and open
symbols. The evolutionary curve for [Cr/Fe] (black trace) was computed with the
chemical evolution model for the solar neighborhood. See text.}
\label{fig1}
\end{center}
\end{figure}

A very low efficiency of inelastic H I collisions is also favored by the
ionization equilbrium of Cr I/Cr II in the metal-poor stars. We have analyzed
$10$ stars\footnote{The details on observational material and derivation of
stellar parameters can be found in Bergemann \& Cescutti (2010).} from different
Galactic populations with spectra obtained by T. Gehren and collaborators with
UVES spectrograph at the VLT (Paranal) and/or with the FOCES spectrograph
at the 2.2m telescope of the CAHA observatory (Calar Alto).
The NLTE Cr I -based abundances in metal-poor stars are systematically larger
than those computed under LTE approach (Fig. 1). The difference of $0.2 - 0.4$
dex is due to substantial overionization of Cr I at low metallicity. The LTE
abundances determined in this work using Cr I lines are consistent with other
LTE studies, confirming that declining [Cr/Fe] with metallicity is an artifact
of the LTE assumption in line formation calculations. The mean NLTE [Cr/Fe]
ratio in stars with subsolar metallicity computed from Cr I lines assuming
$S_{\rm H} = 0$ is $\langle$[Cr/Fe]$\rangle$ $= 0$ with the standard deviation
$\sigma = 0.06$ dex.
Using the Cr II lines, we derive $\langle$[Cr/Fe]$\rangle = -0.05 \pm 0.04$ dex.
The finding that [Cr/Fe] remains constant down to lowest metallicities is
consistent with nucleosynthesis theory, which predicts that Cr and Fe are
co-produced in explosive Si-burning in supernovae in roughly solar proportions.

The NLTE [Cr/Fe] trend with [Fe/H] is reproduced by most of the Galactic
chemical evolution models, without the need to invoke peculiar conditions in the
ISM or to adjust theoretical stellar yields.  The theoretical evolution of
[Cr/Fe] in the solar neighborhood (see Bergemann \& Cescutti 2010 for details),
computed with the two-infall GCE model of Chiappini et al. (1997) is in
agreement with the NLTE results (Fig. 1). The model predicts that $\sim 60$\% of
the total solar Cr and Fe are due to SNe Ia and the rest due to SNe II, the
latter synthesize both elements in roughly solar proportions. The
underproduction of Cr relative to Fe in SNe Ia is compensated by its
overproduction in solar-metallicity SNe II.

\end{document}